\documentclass[10pt,twocolumn,journal]{IEEEtran}
\usepackage{cite}
\usepackage{amsmath,amsfonts,amssymb}
\usepackage{graphicx, epsfig}
\usepackage{array}
\usepackage{multirow}
\usepackage{blkarray}
\usepackage{epsfig,epsf,times,amssymb,amsmath}
\usepackage{algorithmic}
\usepackage{algorithm}

\title{Study of Structured Root-LDPC Codes and PEG Techniques for Block-Fading Channels}

\author{C. T. Healy and Rodrigo C. de Lamare
\thanks{C. T. Healy and R. C. de Lamare are with  CETUC-PUC-Rio, 22453-900 Rio de Janeiro, Brazil, and
also with the Communications Research Group, Department of Electronics,
University of York, YO10 5DD York, U.K. (e-mail: rcdl500@ohm.york.ac.uk). This
work was supported in part by CNPq and FAPERJ in Brazil.}}

\begin{document}
\maketitle

\begin{abstract} 
In this work, we propose structured Root-Low-Density Parity-Check (LDPC) codes
and design techniques for block-fading channels. In particular, Quasi-Cyclic
Root-LDPC codes, Irregular repeat-accumulate Root-LDPC codes and Controlled
Doping Root-LDPC codes based on Progressive Edge Growth (PEG) techniques for
block-fading channels are proposed. The proposed Root-LDPC codes are both
suitable for channels under $F = 2, 3$ and $4$ independent fading per codeword.
The performance of the proposed codes is investigated in terms of Frame Error
Rate (FER). The proposed Root-LDPC codes are capable of achieving the channel
diversity and outperform standard LDPC codes. For block-fading channel with $F
= 2$ our proposed PEG-based Root-LDPC codes outperform PEG-based LDPC codes by
$7.5$dB at a FER close to $10^{-3}$.
\end{abstract}

\section{Introduction}

The most recent IEEE Wireless Local Area Network (WLAN) 802.11ad standard
\cite{wlan.80211ad} argues that to achieve high throughput the devices must
operate with LDPC codes \cite{Gallager, Luby}. As wireless systems are subject
to multi-path propagation and mobility, these systems are characterized by
time-varying channels with fluctuating signal strength. In applications subject
to delay constraints and slowly-varying channels, only limited independent
fading realizations are experienced. In such conditions also known as
non-ergodic scenarios, the channel capacity is zero since there is an
irreducible probability, termed outage probability \cite{Biglieri.98}, that the
transmitted data rate is not supported by the channel. A simple and useful
model that captures the essential characteristics of non-ergodic channels is
the block-fading channel \cite{rappaport, Ozarow}. It is especially important
in wireless communications with slow time-frequency hopping (e.g., cellular
networks and wireless local area networks) or multi-carrier modulation using
Orthogonal Frequency Division Multiplexing (OFDM) \cite{boutros.Trans.10}.
Codes designed for block-fading channels are expected to achieve the channel
diversity and to offer excellent coding gains.

\subsection{Prior and Related Works}

A family of LDPC codes called Root-LDPC for block-fading channels with $F = 2$
fading per codeword was proposed in \cite{boutros.Trans.10}. Root-LDPC codes
are able to achieve the maximum diversity of a block-fading channel and have a
performance near the limit of outage when decoded using the Sum Product
Algorithm (SPA). Root-LDPC codes are always designed with code rate $R = 1/F$,
since the Singleton bound determines that this is the highest code rate
possible to obtain the maximum diversity order \cite{boutros.Trans.10}. Y. Li
and M. Salehi in \cite{salehi.10} have presented the construction of structured
Root-LDPC codes by means of tiling circulant matrices, i.e., by designing
Quasi-Cyclic Low-Density Parity-Check (QC-LDPC) codes \cite{Fossorier}. It is
also shown that the QC-LDPC codes can perform as well as randomly generated
Root-LDPC codes over block-fading channels. Uchoa et.al. in \cite{peg.comms.11}
proposed a PEG-based algorithm to design LDPC codes with root-check properties,
thus providing Root-LDPC codes with larger girths. A strategy that imposes
constraints on a PEG-based algorithm which are required by Root-LDPC codes was
devised. This approach has provided better performance in terms of FER and BER
than the works in \cite{boutros.Trans.10, salehi.10,dopeg}. Duyck et. al. in
\cite{dieter.fulldiv.11} proposed the design of a random LDPC codes which are
able to achieve full diversity in block-fading channels with $F = 2$ fadings.
Healy and de Lamare in \cite{con.iswcs.13} extended the work in
\cite{dieter.fulldiv.11} for the case of block-fading channels with $F = 3$ and
$F = 4$ fading per block transmitted. Uchoa et.al. in \cite{idd.BF.TVT}
proposed iterative detection and decoding (IDD) algorithms for Multiple-Input
Multiple-Output (MIMO) systems operating in block fading and fast Rayleigh
fading channels.

\subsection{Contributions}

We propose in this work three structures to design Root-LDPC codes which are:
Quasi-Cyclic, Repeat and Accumulate and Controlled Doping. Preliminary results
toward PEG-based algorithm to design QC-LDPC codes with root-check properties
for block-fading channel with $F = 3, 4$ fading per codeword were reported in
\cite{iswcs.12}. Here, in this work we present a more detailed analysis of
Quasi-Cyclic root-check based LDPC codes. Furthermore, initial results for a
PEG-based algorithm to design irregular repeat-accumulate (IRA) LDPC codes with
root-check properties for block-fading channels were discussed in
\cite{iswcs.13}. Here, we present a more detailed analysis of Irregular
Repeat-Accumulate and Accumulate IRAA root-check structure for $F = 2, 3$
independent fading.

In general, the parity check bits of Root-LDPC codes are not full diversity.
Boutros in \cite{doping.boutros.11} proposed a controlled doping via high order
Root-LDPC codes, which are able to guarantee full diversity for the parity
check bits. Such a design becomes really important when Iterative Detection and
Decoding (IDD) is used in spread spectrum \cite{poor.99,
rodrigo.DF.CDMA,itic,jidf,jpais,sint} and MIMO systems \cite{mmimo,vblast,
rfa.iet.11,
rodrigo.DF.MIMO,tds,mfsic,dfcc,jiomimo,jpais,armo,ccdfmu,mbthp,did}. In IDD
systems the detector and the decoder exchange their extrinsic information in an
iterative way. Therefore, if the parity bits are not full diversity the overall
IDD system performance will lead to a degradation in terms of Bit Error Rate
(BER) instead of improvements as stated in \cite{poor.99}.

In this paper we also propose a novel full diversity controlled doping
root-check RA-based LDPC codes for Block-Fading channels of $F = 2, 3, 4$
fading which includes the code rates $R = \frac{1}{2}$, $R = \frac{1}{3}$ and
$R = \frac{1}{4}$.

The main contributions of this work can be summarized as:
\begin{itemize}
\item Root-LDPC codes for Block-Fading channels including structured, unstructured, controlled doping, and RA designs are developed.
\item New PEG-based algorithms for several Root-LDPC code structures are presented.
\item A comprehensive simulation study of Root-LDPC codes and design algorithms is detailed.
\end{itemize}

The rest of this paper is organized as follows. In Section 2 we describe the system model. In Section 3 we discuss the prior and related works on the design of Root-LDPC codes and their structure. In Section 4 the proposed PEG-based Quasi-Cyclic Root-LDPC codes, Irregular repeat-accumulate Root-LDPC codes and Controlled Doping Root-LDPC codes and their structure are presented. In Section 5 a discussion of which Root-LDPC code is more appropriate for a specific scenario is provided. Section 6 the simulation results are shown, while Section 7 concludes the paper.

\section{System Model}
Consider a block fading channel, where $F$ is the number of independent fading blocks per codeword of length $N$. Following \cite{salehi.10}, the {\it t}-th received symbol is given by:
\begin{equation} \label{eq:recsymbol}
 r_{t} = h_{f}s_{t}+n_{g_{t}},
\end{equation}
where $1\leq t \leq N$, $1 \leq f \leq F$, $f$ and $t$ are related by $f=\lceil F\frac{t}{N}\rceil$, where $\lceil \phi \rceil$ returns the smallest integer not smaller than $\phi$, $h_f$ is the real Rayleigh fading coefficient of the $f$-th block, $s_{t}$ is the transmitted signal, and $n_{g_{t}}$ is additive white Gaussian noise with zero mean and variance $N_0/2$. In this paper, we assume that the transmitted symbols $s_{t}$ are binary phase shift keying
(BPSK) modulated. We assume that the receiver has perfect channel state information, and that the SNR is defined as $E_b/N_0$, where $E_b$ is the energy per information bit. The information transmission rate is $R=K/N$, where $K$ is the number of information bits per codeword of length $N$. For the case of a block-fading channel, we consider $R=1/F$, since then it is possible to design a practical diversity achieving code \cite{salehi.10}. The performance of a communication system in a non-ergodic block-fading channel can be investigated by means of the outage probability \cite{Biglieri.98},
which is defined as:
\begin{equation} \label{eq:pout}
P_{out} = {\cal P}({\cal I} < R ),
\end{equation}where ${\cal P}(\phi)$ is the probability of event $\phi$ and $\cal I$ is the mutual information. The mutual information $I_{G}$, for Gaussian channel inputs is\cite{salehi.10}:
\begin{equation}\label{eq:outage}
I_{G}=\frac{1}{F} \sum_{f=1}^F \frac{1}{2}\log_2 \left( 1 + 2R\frac{E_b}{N_0}h_f^2\right),
\end{equation}so that an outage occurs when the average mutual information among blocks is smaller than the attempted information transmission rate.

\section{Root-LDPC Codes}

Root-LDPC codes are those which use the graph structure comprising special root-check nodes to ensure full diversity on the block fading channel with greatest possible code rate. These root-checks offer connection from each information node in the graph to the parity bits affected by fading coefficients distinct from that affecting the information node in question. Thus, the information node can be recovered provided at least one fading coefficient is large enough. Since for each information node there is a root-check node for all other fading coefficients, the root-checks appear as identity matrices in the parity-check matrix of the Root-LDPC codes. The properties offered by the root-check node structure are full single-iteration convergence on the noise-free block binary erasure channel and thus full diversity performance on the block fading channel of (\ref{eq:recsymbol}) \cite{boutros.Trans.10}.

In this section, the parity check matrix of the most relevant Root-LDPC codes are discussed. The number of fadings considered are $F = 2, 3$ and $4$ which correspond to code rates $R = \frac{1}{2}, \frac{1}{3}$ and $\frac{1}{4}$.

\subsection{Random Root-LDPC Codes}
Here, we will introduce some definitions and the notation adopted in this work. The binary LDPC code in systematic form is specified by its parity-check matrix \textbf{H}:
\begin{equation}
\mathbf{H} = [\mathbf{I}_{N-K}\; \mathbf{P}],
\label{eq:pcm}
\end{equation}
where $\mathbf{I}_{N-K}$ is the identity matrix of size (N-K) and \textbf{P} is an (N-K)-by-K matrix. Then the generator matrix for the code is:
\begin{equation}
\mathbf{G} = [\mathbf{P}^{T}\; \mathbf{I}_K],
\label{eq:gen}
\end{equation} where $(\cdot)^{T}$ refers to the transpose operation.

The variable node degree sequence $D_s$ is defined to be the set of column weights of \textbf{H} as designed, and is prescribed by the variable node degree distribution $\lambda(x)$ as described in \cite{richardson.01}. Moreover, $D_s$ is arranged in non-decreasing order. The first proposed Root-LDPC codes were devised by Boutros et. al. in \cite{boutros.Trans.10}. Therefore, the general structure of the parity-check matrix for a random Root-LDPC code for $F = 2$ can be defined as
\begin{equation} \label{eq:pcm_random_rc}
\mathbf{H} =
\bordermatrix{
& 1i & 2i & 1p & 2p \cr
1c & \mathbf{I} & \mathbf{H}_{2i} & \mathbf{0} & \mathbf{H}_{2p} \cr
2c & \mathbf{H}_{1i} & \mathbf{I} & \mathbf{H}_{1p} & \mathbf{0} \cr
},
\end{equation}
where the nodes (1i and 2i) represent the information symbols that are sent over two independent fading, the same happens to nodes (1p and 2p) which are the parity symbols; (1c and 2c) are the check nodes. In the parity-check matrix ${\bf H}$, there are eight sub-matrices of size $\frac{N}{4} \times \frac{N}{4}$. \textbf{I} is an identity sub-matrix, {\bf 0} is a null sub-matrix, $\mathit{\mathbf{H}}_{1i}$ and $\mathit{\mathbf{H}}_{2i}$ are sub-matrices of Hamming weight 2 connected to the information symbols, $\mathit{\mathbf{H}}_{1p}$ and $\mathit{\mathbf{H}}_{2p}$ are also sub-matrices of Hamming weight 3 connected to the parity symbols.  In a similar fashion, it can be devised for the case of $F = 3$ as stated in \cite{boutros.Trans.10}.

\subsection{Quasi-Cyclic Root-LDPC Codes}
Following the idea of Boutros et. al. in \cite{boutros.Trans.10}, Li and Salehi in \cite{salehi.10} devised a Quasi-Cyclic Root-LDPC Codes. The parity-check matrix ${\bf H}$ of a QC-LDPC code can be defined as \cite{qcpeg.04}:
\begin{equation}
\mathbf{H} =
\begin{bmatrix}
\mathbf{H}_{0,0} & \mathbf{H}_{0,1} & \cdots & \mathbf{H}_{0,w-1} \\
\mathbf{H}_{1,0} & \mathbf{H}_{1,1} & \cdots & \mathbf{H}_{0,w-1} \\
\vdots & \vdots & \ddots & \vdots \\
\mathbf{H}_{c-1,0} & \mathbf{H}_{c-1,1} & \cdots & \mathbf{H}_{c-1,w-1} \\
\end{bmatrix},
\label{eq:qcmatric}
\end{equation}
where ${\bf H}_{ij}$ is an $n\times n$ circulant or all-zeros matrix, and $c$ and $w$ are two positive integers with $c<w$. The null space of ${\bf H}$ gives a QC-LDPC code over $GF(2)$ of length $N = wn$. The rank of ${\bf H}$ is at most $cn$. Hence the code rate is at least $\frac{w-c}{w}$.

For the case  of Quasi-Cyclic Root-LDPC codes the parity-check matrix follows the same idea as (\ref{eq:pcm_random_rc}), although the sub-matrices become a set of Quasi-Cyclic matrices. Consequently, $\mathbf{I}$ becomes
\begin{equation} \label{eq:i_pcm}
\mathbf{I}_{(top-left)} = \begin{vmatrix}
 \mathbf{I}_{0,0} & \mathbf{0} & \mathbf{0} & \mathbf{0} \\
 \mathbf{0} & \mathbf{I}_{1,1} & \mathbf{0} & \mathbf{0} \\
 \mathbf{0} & \mathbf{0} & \mathbf{I}_{2,2} & \mathbf{0} \\
  \mathbf{0} & \mathbf{0} & \mathbf{0} & \mathbf{I}_{3,3}
\end{vmatrix},
\end{equation} $\mathbf{H}_{1i}$ as
\begin{equation} \label{eq:h1i_pcm}
\mathbf{H}_{1i} = \begin{vmatrix}
 \mathbf{I}_{4,0} & \mathbf{I}_{4,1} & \mathbf{0} & \mathbf{0} \\
 \mathbf{0} & \mathbf{I}_{5,1} & \mathbf{I}_{5,2} & \mathbf{0} \\
 \mathbf{0} & \mathbf{0} & \mathbf{I}_{6,2} & \mathbf{I}_{6,3} \\
  \mathbf{I}_{7,0} & \mathbf{0} & \mathbf{0} & \mathbf{I}_{7,3}
\end{vmatrix}
\end{equation} and for $\mathbf{H}_{1p}$ we define it as
\begin{equation} \label{eq:h1p_pcm}
\mathbf{H}_{1p} = \begin{vmatrix}
 \mathbf{0} & \mathbf{I}_{4,5} & \mathbf{I}_{4,6} & \mathbf{I}_{4,7} \\
 \mathbf{I}_{5,4} & \mathbf{0} & \mathbf{I}_{5,6} & \mathbf{I}_{5,7} \\
 \mathbf{I}_{6,4} & \mathbf{I}_{6,5} & \mathbf{0} & \mathbf{I}_{6,7} \\
  \mathbf{I}_{7,4} & \mathbf{I}_{7,5} & \mathbf{I}_{7,6} & \mathbf{0}
\end{vmatrix},
\end{equation}
where each $\mathbf{I}_{i,j}$ is a circulant permutation matrix, a circulant matrix with row and column weights $1$. Each $\mathbf{0}$ is a null matrix. The matrix $\mathbf{H}_{2i}$ is similarly formed of tiled circulant  permutation matrices with random cyclic shift, and constrained random placement of the non-null matrices to achieve the required column and row weights. The matrix $\mathbf{H}_{2p}$ has the same form as (\ref{eq:h1p_pcm}) in order that the parity part of the matrix has full rank, but with distinct random cyclic shifts \cite{salehi.10}. The example presented in Equations (\ref{eq:i_pcm}), (\ref{eq:h1i_pcm}) and (\ref{eq:h1p_pcm}) are for a regular QC-Root-LDPC code $C(3,6)$. QC-Root-LDPC codes were proposed with the aim of providing fast encoding and to save memory to store the generator matrix. Li and Salehi in \cite{salehi.10} have shown that the QC-LDPC codes can perform as well as randomly generated Root-LDPC codes \cite{boutros.Trans.10} over block-fading channels.

\subsection{Unstructured Full Diversity LDPC Codes}
Duyck et. al. in \cite{dieter.fulldiv.11} proposed the design of random LDPC codes which are able to achieve full diversity in block-fading channels with $F = 2$ fading. The principle proposed in \cite{dieter.fulldiv.11} is to allow a small reduction in coding rate in order to produce random codes that may achieve the diversity of the channel, i.e., the error rate achieved by the code behaves as $\frac{1}{SNR^2}$. However, as these codes achieve the desired error rate performance but do not have the maximal rate allowed by the Singleton bound, they may be called full diversity codes but not blockwise maximum-distance separable (MDS) codes \cite{Guillen}. Specifically, the codes of \cite{dieter.fulldiv.11} place the requirements that the nodes associated with the information bits have weight $d_{v} = 2$ and do not participate in any stopping sets. The code rate is $R \cong 0.5$.


The design of such LDPC codes was achieved by requiring that the number of check nodes in the graph be greater than $\frac{N}{2}$, i.e., that the rate be less than $\frac{1}{2}$, and that the weight of the first $\frac{N}{2}$ variable nodes is $2$ and that the graph be constructed by the PEG algorithm \cite{hu.05}, which maximises cycle length at each placement, ensuring under these conditions no cycles in the sub-graph comprised of the first $\frac{N}{2}$ variable nodes alone. The requirement of recoverability for the worst-case scenario is equivalent to the requirement that no information variable node $v_{inf} \in \mathbf{V}_{inf}$, affected by $\alpha_1$, is an element of any stopping set found among the variable nodes $\mathbf{V}_1 \cup \{ \mathbf{V}_2 \cup \mathbf{V}_3 \cup \cdots \mathbf{V}_F \}\backslash \mathbf{V}_i$. This requirement must hold for all $i = 2, \cdots, F$ for the information variable nodes to be recoverable on the block binary erasure channel and thus for the code to achieve full diversity on the block fading channel. The parity-check matrix for this general case, with variable node subset labels and the corresponding fading coefficients are given in Fig. 1.

\subsubsection{Unstructured Full Diversity Rate $\frac{1}{3}$}
In (\ref{eq:H_BF3_soln}) is shown a code graph for the case of $F = 3$ fading per codeword \cite{con.iswcs.13} by means of imposing null matrices on the parity-check matrix, along with restrictions on the cycles present in the sub-graphs of the code. The structured matrices $[\mathbf{H}_{\alpha,1} \mathbf{H}_{\alpha_{2}}]$ and $[\mathbf{H}_{\alpha,2} \mathbf{H}_{\alpha_{3}}]$ must be constructed by the PEG algorithm, as in \cite{dieter.fulldiv.11}, ensuring the extrinsic connections to $\mathbf{V}_2$ and $\mathbf{V}_3$,  respectively. The constraints on the code sub-graphs result in the variable nodes of $\mathbf{V}_1$ having weight $4$. The distribution of the nodes in $\mathbf{V}_2$ and $\mathbf{V}_3$ is unconstrained and may be irregular. In addition to this weight constraint, each of the sub-matrices $[\mathbf{H}_{\alpha,1} \mathbf{H}_{\alpha_{2}}]$ and $[\mathbf{H}_{\alpha,2} \mathbf{H}_{\alpha_{3}}]$ are constrained to have rate less than $\frac{1}{2}$, and so the final graph will have rate less than $\frac{1}{3}$.

\begin{equation} \label{eq:H_BF3_soln}
\mathbf{H}_{BF3} =
\begin{blockarray}{ccc}
\alpha_{1} & \alpha_{2} & \alpha_{3} \\
\begin{block}{[ccc]}
  \mathbf{H}_{\alpha_{1},1} & \mathbf{H}_{\alpha_{2}} & \mathbf{0} \\
  \mathbf{H}_{\alpha_{1},2} & \mathbf{0} & \mathbf{H}_{\alpha_{3}} \\
\end{block}
\end{blockarray}
\end{equation}


\subsubsection{Unstructured Full Diversity Rate $\frac{1}{4}$}
The code graph achieving the requirements on stopping sets among $\mathbf{V}_1, \cdots, \mathbf{V}_4$ containing information variable nodes is presented in (\ref{eq:H_BF4_soln}) \cite{con.iswcs.13}. We can see that with each additional fading coefficient considered, a straightforward graph expansion is carried out, effectively nesting the $F-1$ diversity achieving graph in the code capable of full diversity performance on the channel with $F$ fading coefficients.

\begin{equation} \label{eq:H_BF4_soln}
\mathbf{H}_{BF4} =
\begin{blockarray}{cccc}
\alpha_{1} & \alpha_{2} & \alpha_{3} & \alpha_{4} \\
\begin{block}{[cccc]}
  \mathbf{H}_{\alpha_{1},1} & \mathbf{H}_{\alpha_{2}} & \mathbf{0} & \mathbf{0} \\
  \mathbf{H}_{\alpha_{1},2} & \mathbf{0} & \mathbf{H}_{\alpha_{3}} & \mathbf{0} \\
  \mathbf{H}_{\alpha_{1},3} & \mathbf{0} & \mathbf{0} & \mathbf{H}_{\alpha_{4}} \\
\end{block}
\end{blockarray}
\end{equation}

%

\section{Proposed PEG-Based Root-LDPC Codes}
In this section, the proposed PEG-Based Root-LDPC codes are discussed. The number of fadings considered are $F = 2, 3$ and $4$ which correspond to code rates $R = \frac{1}{2}, \frac{1}{3}$ and $\frac{1}{4}$.

\subsection{QC PEG-Based Root-LDPC Codes}
Preliminary results on the design of a PEG-based Quasi-
Cyclic Root-LDPC codes for Block-Fading channel with $F = 3, 4$ fadings per codeword were presented by Uchoa et. al. in \cite{iswcs.12}. The codes generated by this strategy can achieve a significant performance in terms of FER with respect to the theoretical limit. These codes can save up to $3$dB in terms of signal to noise ratio to achieve the same FER when compared to other codes.

A Root-LDPC code requires a designer to divide both variable and check nodes in $F$ equal parts. Following the root-check based structure reported in \cite{boutros.Trans.10}, the parity-check matrix becomes:
\begin{equation}
{\bf H} = [\mathbf{S}_{1}\mathbf{P}_{1}, \cdots, \mathbf{S}_{F}\mathbf{P}_{F}],
\label{eq:hrcf3}
\end{equation}
where the subscripts represent the variable nodes (information and parity, respectively) under a specific fading block. The parity-check matrix of (\ref{eq:hrcf3}) can be reordered to $\mathbf{H} = [\mathbf{S}_{1}, \cdots, \mathbf{S}_{F}\mathbf{P}_{1}, \cdots, \mathbf{P}_{F}]$, with the blocks $\mathbf{S}_i$ associated with information nodes and the blocks $\mathbf{P}_i$ associated with parity nodes. In order to obtain the generator matrix, the sub-matrix ${\bf B}$ formed by parity matrices $\mathbf{P}_{1}, \cdots, \mathbf{P}_{F}$ must be a non-singular matrix, which means it is invertible under $GF(2)$ \cite{salehi.10}.

To design a practical code for $F = 3$ which is able to achieve the channel diversity, the highest possible rate of such a code is $R = \frac{1}{F}=\frac{1}{3}$.  As a result, the parity-check matrix for $R=\frac{1}{3}$ can be defined as in (\ref{eq:pcmsysf3}),

\begin{equation}
\mathbf{H}=
\begin{blockarray}{cccccc}
1i & 2i & 3i & 1p & 2p & 3p\\
\begin{block}{[cccccc]}
\begin{array}{c}\mathbf{I_{0,0}}\\ \mathbf{I_{1,0}}\\ \mathbf{H_{2,0}}\\ \mathbf{0} \\ \mathbf{H_{4,0}} \\ \mathbf{0} \\ \end{array} & \begin{array}{c}\mathbf{H_{0,1}} \\ \mathbf{0} \\ \mathbf{I_{2,1}} \\ \mathbf{I_{3,1}} \\ \mathbf{0} \\ \mathbf{H_{5,1}} \\\end{array} & \begin{array}{c} \mathbf{0}\\ \mathbf{H_{1,2}} \\ \mathbf{0} \\ \mathbf{H_{3,2}} \\ \mathbf{I_{4,2}} \\ \mathbf{I_{5,2}}  \end{array} & \begin{array}{cc}\mathbf{0} & \mathbf{0}\\\mathbf{0} & \mathbf{0}\\ \mathbf{0} & \mathbf{0}\\\mathbf{H_{3,3}} & \mathbf{0}\\\mathbf{H_{4,3}} & \mathbf{H_{4,4}}\\\mathbf{H_{5,3}} & \mathbf{H_{5,4}} \end{array} & \begin{array}{cc}\mathbf{0} & \mathbf{H_{0,6}}\\\mathbf{0} & \mathbf{0}\\\mathbf{0} & \mathbf{0}\\\mathbf{0} & \mathbf{0}\\\mathbf{0} & \mathbf{0}\\\mathbf{H_{5,5}} & \mathbf{0} \end{array} & \begin{array}{cc}\mathbf{H_{0,7}} & \mathbf{H_{0,8}}\\\mathbf{H_{1,7}} & \mathbf{H_{1,8}}\\\mathbf{0} & \mathbf{H_{2,8}} \\\mathbf{0} & \mathbf{0} \\\mathbf{0} & \mathbf{0} \\\mathbf{0} & \mathbf{0}  \end{array}\\
\end{block}
\end{blockarray},
\label{eq:pcmsysf3}
\end{equation}


where the $n\times n$ matrices ${\bf H}_{ij}$ are circulant matrices of column and row weight as required by the degree distribution of the code, ${\bf I}_{ij}$ are $n\times n$ circulant permutation matrices, while $\mathbf{0}$ is an all-zeros matrix. The notation ${\bf I}_{ij}$ was used to reinforce that such connections are the root-check connections \cite{boutros.Trans.10}. The restrictions that should be imposed are only the ${\bf I}_{ij}$ to be placed in the positions described in (\ref{eq:pcmsysf3}) and the upper and down triangular sub-matrices in the parity part, ${\bf B}$, of ${\bf H}$. In order to perform a PEG-based design the only restriction imposed is that the sub-matrices ${\bf I}_{ij}$ and the upper and down sub-matrices of (\ref{eq:pcmsysf3}) are kept. The other sub-matrices can be placed following a quasi-cyclic PEG-based algorithm.

The parity-check matrix for $F = 4$ with code rate $R=\frac{1}{4}$ is structured similarly to (\ref{eq:pcmsysf3}), and the same restrictions may be imposed to the design to construct a PEG-based QC-Root-LDPC code for this scenario.

\subsubsection{Proposed Design Algorithm}
Here, we introduce some definitions and notations. Then, we present the pseudo-code of our proposed algorithm for PEG-based Quasi-Cyclic Root-LDPC codes. The block-fading channels with $F = 3$ and $F = 4$ are considered. In extending to a greater number of fadings, $F>4$, the general structure presented is maintained, with the information variable nodes for each fading possessing root-check identity matrices connecting to parity variable nodes in each of the other fading blocks only, ensuring the upper and lower triangular sections of parity bits observed in (\ref{eq:pcmsysf3}). The placement of the remaining cyclic sub-matrices is required to maintain this relationship and provide satisfactory final code degree distribution. The LDPC code is specified by its sparse parity-check matrix $\mathbf{H} = [\mathbf{A}~\vert~\mathbf{B}]$, where $\mathbf{A}$ is a matrix of size $M$-by-$K$, and $\mathbf{B}$ is an $M$-by-$M$ matrix. The generator matrix for the code is $\mathbf{G} = [(\mathbf{B}^{-1}\mathbf{A})^{T}~\vert~ \mathbf{I}_K]$, $\mathbf{I}_{K}$ is an identity matrix of size $K$.

The variable node degree sequence $D_s$ is defined as the set of column weights of the designed \textbf{H}, and is prescribed by the variable node degree distribution $\lambda(x)$ as described in \cite{richardson.01}. Moreover, $D_s$ is arranged in non-decreasing order. The proposed algorithm, called QC-PEG Root-LDPC, constructs \textbf{H} by operating progressively on variable nodes to place the edges required by $D_s$. The Variable Node of interest is labelled $v_j$ and the candidate check nodes are individually referred to as $c_i$. The PEG Root-LDPC algorithm chooses a check node $c_i$ to connect to the variable node of interest $v_j$ by expanding a constrained sub-graph from $v_j$ up to maximum depth $l$. The set of check nodes found in this sub-graph are denoted $N_{v_j}^l$ while the set of check nodes of interest, those not currently found in the sub-graph, are denoted $\overline{N_{v_j}^l}$. For the QC-PEG Root-LDPC algorithm, a check node is chosen at random from the minimum weight check nodes of this set.

To impose the Root-LDPC structure it is necessary simply to initialize the graph with root-check connections, which appear as the identity matrices in the parity-check matrix of the code, and to ensure no additional edge placement is made either in the identity matrices or the null matrices specified by the Root-LDPC structure. This is achieved in the PEG algorithm by modification of the indicator vector presented in \cite{peg.comms.11}. Zeros in the indicator vectors, as presented in the following section, exclude check nodes from the expanded tree of the PEG algorithms and this exclude edge placement connecting to those check nodes.

\subsubsection{Pseudo-code for the QC-PEG-Root-LDPC Algorithm}
Initialization: A matrix of size $M \times N$ is created with the circulant permutation matrices $\mathbf{I}_{i,j}$ in the positions shown in (\ref{eq:pcmsysf3}) and zeros in all other positions. We define the indicator vectors ${\mathbf{z}}_{1}, \cdots, {\mathbf{z}}_{F^2}$ for the $R = \frac{1}{3}$ case as:

\begin{eqnarray}
{\mathbf{z}}_{1} & = & [\mathbf{0}_{1 \times \frac{2N}{9}}, \mathbf{1}_{1 \times \frac{N}{9}}, \mathbf{0}_{1 \times \frac{N}{9}}, \mathbf{1}_{1 \times \frac{N}{9}}, \mathbf{0}_{1 \times \frac{N}{9}}]^{T},\nonumber\\
{\mathbf{z}}_{2} & = & [\mathbf{1}_{1 \times \frac{N}{9}}, \mathbf{0}_{1 \times \frac{4N}{9}}, \mathbf{1}_{1 \times \frac{N}{9}}]^{T},\nonumber\\
{\mathbf{z}}_{3} & = & [\mathbf{0}_{1 \times \frac{N}{9}}, \mathbf{1}_{1 \times \frac{N}{9}}, \mathbf{0}_{1 \times \frac{N}{9}}, \mathbf{1}_{1 \times \frac{N}{9}}, \mathbf{0}_{1 \times \frac{2N}{9}}]^{T},\nonumber\\
{\mathbf{z}}_{\chi} & = & [\mathbf{0}_{1 \times \frac{(i-1) \cdot N}{9}}, \mathbf{1}_{1 \times \frac{(7-i)N}{9}}]^{T}~~ \rm{for} ~ \chi = 4,5,6, \nonumber\\
{\mathbf{z}}_{\gamma} & = & [\mathbf{1}_{1 \times \frac{(i-6) \cdot N}{9}}, \mathbf{0}_{1 \times \frac{(12-i)N}{9}}]^{T}~~ \rm{for} ~ \gamma = 7,8,9, \nonumber\\
\label{eq:indf3}
\end{eqnarray}

The indicator vectors for the construction of the QC-PEG-Root-LDPC code with $R = \frac{1}{4}$ designed similarly to (\ref{eq:pcmsysf3}) but for the channel with $F=4$ are:

\begin{eqnarray}
{\mathbf{z}}_{1} & = & [\mathbf{0}_{1 \times \frac{3N}{16}}, \mathbf{1}_{1 \times \frac{N}{16}}, \mathbf{0}_{1 \times \frac{N}{16}}, \mathbf{1}_{1 \times \frac{2N}{16}},  \mathbf{0}_{1 \times \frac{2N}{16}}, \nonumber\\
& & ~~~ \mathbf{1}_{1 \times \frac{N}{16}}, \mathbf{0}_{1 \times \frac{N}{16}}, \mathbf{1}_{1 \times \frac{N}{16}}]^T,\nonumber\\
{\mathbf{z}}_{2} & = & [\mathbf{1}_{1 \times \frac{N}{16}}, \mathbf{0}_{1 \times \frac{N}{16}}, \mathbf{1}_{1 \times \frac{N}{16}}, \mathbf{0}_{1 \times \frac{4N}{16}}, \mathbf{1}_{1 \times \frac{N}{16}}, \nonumber\\
& & ~~~  \mathbf{0}_{1 \times \frac{2N}{16}}, \mathbf{1}_{1 \times \frac{2N}{16}},]^T,\nonumber\\
{\mathbf{z}}_{3} & = & [\mathbf{0}_{1 \times \frac{N}{16}}, \mathbf{1}_{1 \times \frac{N}{16}}, \mathbf{0}_{1 \times \frac{N}{16}}, \mathbf{1}_{1 \times \frac{2N}{16}} , \mathbf{0}_{1 \times \frac{4N}{16}},  \nonumber\\
& & ~~~   \mathbf{1}_{1 \times \frac{2N}{16}}, \mathbf{0}_{1 \times \frac{N}{16}}]^T,\nonumber\\
{\mathbf{z}}_{4} & = & [\mathbf{1}_{1 \times \frac{N}{16}}, \mathbf{0}_{1 \times \frac{N}{16}}, \mathbf{1}_{1 \times \frac{2N}{16}}, \mathbf{0}_{1 \times \frac{N}{16}}, \mathbf{1}_{1 \times \frac{N}{16}},  \nonumber\\
& & ~~~   \mathbf{0}_{1 \times \frac{2N}{16}}, \mathbf{1}_{1 \times \frac{N}{16}}, \mathbf{0}_{1 \times \frac{3N}{16}},]^T,\nonumber\\
{\mathbf{z}}_{\chi} & = & [\mathbf{0}_{1 \times \frac{(i+1)N}{16}}, \mathbf{v}_{ALT}{\scriptstyle(0:\frac{(11-i)N}{16} - 1)}]^T\nonumber\\ & & ~~~ \rm{for} ~ \chi = 5,\cdots,10,\nonumber\\
{\mathbf{z}}_{\gamma} & = & [\mathbf{v}_{ALT}{\scriptstyle(\frac{(17-i)N}{16}:\frac{7N}{16}-1)}, \mathbf{0}_{1 \times \frac{(22-i)N}{16}}]^T\nonumber\\ & & ~~~ \rm{for} ~ \gamma = 11,\cdots,16,\nonumber\\
\label{eq:indf4}
\end{eqnarray}

\vspace{-.6cm}

\begin{equation}
{\mathbf{v}}_{ALT} = [\mathbf{1}_{1 \times \frac{N}{16}},  \mathbf{0}_{1 \times \frac{N}{16}},\mathbf{1}_{1 \times \frac{N}{16}},  \mathbf{0}_{1 \times \frac{N}{16}},\mathbf{1}_{1 \times \frac{N}{16}},  \mathbf{0}_{1 \times \frac{N}{16}},\mathbf{1}_{1 \times \frac{N}{16}}]
\end{equation}

These indicator vectors are modelled on that of the original PEG algorithm \cite{hu.05}, indicating submatrices for which placement is permitted, thus imposing the required form. The degree sequence as defined for LDPC codes must be altered to take into account the structure imposed by Root-LDPC codes, namely the circulant permutation matrices, $\mathbf{I}_{i,j}$, of (\ref{eq:pcmsysf3}) and similarly the structure defined by (\ref{eq:indf4}). The pseudo-code for our proposed QC-PEG Root-LDPC algorithm is detailed in Algorithm \ref{alg:rcpegalg1}, where the indicator vector, $\mathbf{z}_i$, is taken from (\ref{eq:indf3}), (\ref{eq:indf4}) for constructing codes of rate $R=\frac{1}{3}$, $R=\frac{1}{4}$, respectively.

\begin{algorithm}
 \caption{QC-PEG Root-LDPC Algorithm}
 \label{alg:rcpegalg1}
\algsetup{
linenosize=\small,
linenodelimiter=.
}
\begin{algorithmic}[1]
\FOR{$j = 1 : F^2$}
\FOR{$k = 0 : D_s(j)-1$}
\IF{$j\ge\frac{N}{F}$ ~ \& ~ $k == 0$  }
\STATE Place edge at random among minimum weight submatrices permitted by the indicator ${\mathbf{z}}_{j}$, with a random first edge placement within the chosen submatrix, in column $\frac{(j-1)\cdot N}{F^2}$\emph{-th}.
\STATE Place remaining edges in the submatrix by circulant shift of the first placement.
\STATE Null the entry in the indicator vector ${\mathbf{z}}_{j}$ in the position of the chosen submatrix, preventing further placements in that submatrix.
\ELSE
\STATE Expand the PEG subtree from the $\frac{(j-1)\cdot N}{F^2}$\emph{-th} variable node to depth \emph{l} such that the tree contains all check nodes allowed by the indicator vector \textbf{or} the number of nodes in the tree does not increase with an expansion to the \emph{(l+1)-th} level.
\STATE Place edge connecting the $\frac{(j-1)\cdot N}{F^2}$\emph{-th} variable node to a check node chosen randomly from the set of minimum weight nodes which were added to the subtree at the last tree expansion.
\STATE Place remaining edges in the submatrix by circulant shift of the first placement.
\STATE Null the entry in the indicator vector ${\mathbf{z}}_{j}$ in the position of the chosen submatrix, preventing further placements in that submatrix.\ENDIF
\ENDFOR
\ENDFOR
\end{algorithmic}
\end{algorithm}

\subsection{RA Based Root-LDPC Codes}
Preliminary results on the PEG-based design of Irregular Repeat Accumulate (IRA) LDPC codes \cite{Jin} with root-check properties were reported in \cite{iswcs.13}. We considered a block-fading channel with $F = 2$ and $F = 3$. Here, in this section we synthesize the most relevant information on the design of IRA Root-LDPC codes.

A repeat-accumulate (RA) code consists of a serial concatenation, through an interleaver, of a single rate $1/q$ repetition code with an accumulator having transfer function $\frac{1}{1+D}$, where $q$ is the number of repetitions for each group of $K$ information bits. Fig. 2 shows a typical repeat-accumulate code block diagram. The implementation of the transfer function $\frac{1}{1+D}$ is identical to an accumulator,
although the accumulator value can be only $0$ or $1$ since the operations are over the binary field \cite[pp. 267-279]{ryanbook}. As discussed in \cite[pp. 267-279]{ryanbook}, to ensure a large minimum Hamming distance, the interleaver should be designed so that consecutive 1s at its input are widely separated at its output. The RA based codes proposed in \cite{iswcs.13} were systematic. The main limitation of RA codes on Gaussian channels is the code rate, which cannot be higher than $\frac{1}{2}$. This limitation is not relevant for block-fading channels as the rate is constrained to be $R \leq \frac{1}{2}$ in order to achieve a diversity order greater than or equal to 2.

Irregular repeat-accumulate (IRA) codes generalize the concept of RA codes by changing the repetition rate for each group of $K$ information bits and performing a linear combination of the repeated bits which are sent through the accumulator. Furthermore, IRA codes are typically systematic. IRA codes allow flexibility in the choice of the repetition rate for each information bit so that high-rate codes may be designed. Their irregularity allows
operation closer to the capacity limit \cite[pp. 267-279]{ryanbook}.

The parity-check matrix for a systematic RA and IRA codes has the form $\mathbf{H} = \left[\mathbf{H}_u~\mathbf{H}_p\right]$, where $\mathbf{H}_p$ is a square dual-diagonal matrix given by
\begin{equation}
 \mathbf{H}_p =
    \begin{bmatrix}
    1 &  &  & & \\
    1 & 1 & & & \\
      & \ddots & \ddots & &\\
      & & 1 & 1 & \\
      & &  & 1 & 1\\
    \end{bmatrix}.
\label{eq:pcm_ra_ira}
\end{equation}
For RA codes, $\mathbf{H}_u$ is a regular matrix having column weight $q$ and row weight $1$. For IRA codes, $\mathbf{H}_u$ has irregular columns and rows weights. The Generator Matrix (GM) can be obtained as $\mathbf{G} = \left[\mathbf{I}_{K}~\mathbf{H}_{u}^{T}\mathbf{H}_{p}^{-T}\right]$,
where $\mathbf{I}_{K}$ is an identity matrix of dimension $K \times K$, and the matrix $\mathbf{H}_{p}^{-T}$ is the well-known inverse transpose of (\ref{eq:pcm_ra_ira}).

\subsubsection{IRA Root-LDPC Rate $\frac{1}{2}$}
The design of a Root-LDPC code with an IRA structure imposes some constraints in terms of parity-check matrix to guarantee the root-check properties. Following the notation adopted in \cite{iswcs.11}, for the case of a systematic Rate $\frac{1}{2}$ with $F = 2$, the parity-check matrix must be like
\begin{equation}
 \mathbf{H}=
    \begin{bmatrix}
    \mathbf{I}_{\frac{N}{4}} & \mathbf{H}_{2} & \mathbf{0}_{\frac{N}{4}} & \mathbf{H}_{3}\\
    \mathbf{H}_{2} & \mathbf{I}_{\frac{N}{4}} & \mathbf{H}_{3} & \mathbf{0}_{\frac{N}{4}}\\
    \end{bmatrix},
\label{eq:pcm_rc_r12}
\end{equation}
where $\mathbf{H}_{2}$ and $\mathbf{H}_{3}$ are $\frac{N}{4}\times\frac{N}{4}$ sub-matrices with Hamming weight two and three, respectively, while $\mathbf{0}_{\frac{N}{4}}$ is a null sub-matrix with dimension $\frac{N}{4}\times\frac{N}{4}$. Therefore, to impose the RA structure and root-check properties the parity-check matrix of an IRA Root-LDPC is
\begin{equation}
 \mathbf{H}=
    \begin{bmatrix}
    \mathbf{I}_{\frac{N}{4}} & \mathbf{H}_{2} & \mathbf{0}_{\frac{N}{4}} & \mathbf{H}_{p}\\
    \mathbf{H}_{2} & \mathbf{I}_{\frac{N}{4}} & \mathbf{H}_{p} & \mathbf{0}_{\frac{N}{4}}\\
    \end{bmatrix},
\label{eq:pcm_ira_rc_r12}
\end{equation}
where $\mathbf{H}_{p}$ is a dual diagonal matrix with dimension $\frac{N}{4}\times\frac{N}{4}$.

\subsubsection{IRA Root-LDPC Rate $\frac{1}{3}$}
For the case of Rate $\frac{1}{3}$ with $F = 3$, we followed a similar structure to the one adopted in \cite{iswcs.12, boutros.Trans.10}. The accumulator used is a transfer function given by $\frac{1}{1+D+D^{\frac{N}{9}}}$ as suggested by \cite{sarah.johnson.05} for the Gaussian channel, and used here to improve coding gain by allowing a more complete connection between the parity bits and the root-check identity matrix through $\mathbf{H}_p$. As a result of the root-check structure of the graph where each root-check identity matrix must connect through a matrix of size $\frac{N}{9}\times\frac{2N}{9}$ to a set of parity bits affected by some other fading coefficient, $\mathbf{H}_{p}$ must be redefined as

\begin{equation}
 \mathbf{H}_{p} =
    \begin{bmatrix}
    \mathbf{H}_{p1} \\
    \mathbf{H}_{p2} \\
    \end{bmatrix},
\label{eq:hpira3}
\end{equation}

\begin{equation}
\mathbf{H}_{p1} =
 \begin{bmatrix}
 1 & 0 & \cdots & \cdots & \cdots & \cdots & \cdots & 0\\
 1 & 1 & 0 & \ddots & \ddots & \ddots & \ddots & 0\\
 \vdots & \ddots & \ddots & 0 & \ddots & \ddots & \ddots & 0\\
  0 & 0 & 1 & 1 & 0 & 0 & 0 & 0\\
 \end{bmatrix},
\label{eq:hp1}
\end{equation}

\begin{equation}
\mathbf{H}_{p2} =
 \begin{bmatrix}
  1 & 0 & \cdots & 0 & 1 & 0 & \cdots & 0\\
  0 & \ddots & 0 & 0 & 1 & 1 & 0 & 0\\
  \vdots & 0 & \ddots & 0 & 0 & \ddots & \ddots & 0\\
  0 & 0 & 0 & 1 & 0 & 0 & 1 & 1\\
 \end{bmatrix},
 \label{eq:hp2}
\end{equation}
where $\mathbf{H}_{p1}$ and $\mathbf{H}_{p2}$ are sub-matrices with dimensions $\frac{N}{9}\times\frac{2N}{9}$. Therefore, the parity-check matrix $\mathbf{H} = [\mathbf{H}_{u} \vert \mathbf{H}_{p}]$ for this particular case of an IRA Root-LDPC Rate $\frac{1}{3}$ as
\begin{equation}
 \mathbf{H}=
    \begin{bmatrix}
     \mathbf{I}_{\frac{N}{9}} & \mathbf{H}_{1} & \mathbf{0}_{\frac{N}{9}} & \vert & \mathbf{0} & \mathbf{H}_{p2} & \mathbf{0}\\
     \mathbf{I}_{\frac{N}{9}} & \mathbf{0}_{\frac{N}{9}} & \mathbf{H}_{1} & \vert & \mathbf{0} & \mathbf{0} & \mathbf{H}_{p1}\\
     \mathbf{H}_{1} & \mathbf{I}_{\frac{N}{9}} & \mathbf{0}_{\frac{N}{9}} & \vert & \mathbf{H}_{p1} & \mathbf{0} & \mathbf{0}\\
     \mathbf{0}_{\frac{N}{9}} & \mathbf{I}_{\frac{N}{9}} & \mathbf{H}_{1} & \vert & \mathbf{0} & \mathbf{0} & \mathbf{H}_{p2}\\
     \mathbf{H}_{1} & \mathbf{0}_{\frac{N}{9}} & \mathbf{I}_{\frac{N}{9}} & \vert & \mathbf{H}_{p2} & \mathbf{0} & \mathbf{0}\\
     \mathbf{0}_{\frac{N}{9}} & \mathbf{H}_{1} & \mathbf{I}_{\frac{N}{9}} & \vert & \mathbf{0} & \mathbf{H}_{p1} & \mathbf{0}\\
    \end{bmatrix},
\label{eq:pcm_ira_rc_r13}
\end{equation}
where $\mathbf{H}_{1}$ and $\mathbf{I}_{\frac{N}{9}}$ are sub-matrices with dimensions $\frac{N}{9}\times\frac{N}{9}$ and
$\mathbf{H}_{1}$ is a sub-matrix with Hamming weight equal to $1$. The null sub-matrices $\mathbf{0}$ on the right hand side of
(\ref{eq:pcm_ira_rc_r13}) have dimensions $\frac{N}{9}\times\frac{2N}{9}$ while on the left hand side the dimensions are $\frac{N}{9}\times\frac{N}{9}$.

\subsection{IRAA Root-LDPC Design}
The general structure of an Irregular Repeat-Accumulate and
Accumulate (IRAA) encoder can be seen in Fig. 3. In this figure, some $b$ extra parity bits are indicated in addition to the normal $p$ parity bits. The $b$ parity
bits can be punctured to obtain a higher code rate. For instance, in
general an IRAA code has rate $1/3$ without puncturing, while by
puncturing $b$ parity-checks a code with rate $1/2$ can be obtained.

The parity-check matrix of an IRAA LDPC code can be represented by
\begin{equation}
\mathbf{H} =
\begin{bmatrix}
    \mathbf{H}_{u} & \mathbf{H}_{p} & \mathbf{0}\\
    \mathbf{0} & \prod_{1} & \mathbf{H}_{p}\\
\end{bmatrix},
\label{eq:pcm_iraa}
\end{equation} where $\prod_{1}$ must be a sub-matrix with rows and columns with Hamming weight one.

In order to obtain IRAA Root-LDPC codes some
constraints must be imposed on the standard IRAA design. We have
noticed that the IRAA Root-LDPC codes led to a more flexible
rate compatible code. For further details refer to \cite{iswcs.13}.

\subsubsection{IRAA Root-LDPC Rate $\frac{1}{2}$}
We applied the root-check structure from (\ref{eq:pcm_ira_rc_r12}) in (\ref{eq:pcm_iraa}) to obtain the following parity-check matrix for rate $1/2$
\begin{equation}
\mathbf{H} =
\left[
\begin{array}{cccccc}
\mathbf{I}_{\frac{N}{9}} & \mathbf{H}_{2} & \mathbf{0}_{\frac{N}{9}} & \mathbf{H}_{p} & \mathbf{0} & \mathbf{0}_{\frac{N}{9}}\\
\mathbf{H}_{2} & \mathbf{I}_{\frac{N}{9}} & \mathbf{H}_{p} & \mathbf{0}_{\frac{N}{9}} & \mathbf{0}_{\frac{N}{9}} & \mathbf{0}_{\frac{N}{9}}\\
\mathbf{0}_{\frac{N}{9}} & \mathbf{0}_{\frac{N}{9}} & \multicolumn{2}{c}{\multirow{2}{*}{$\prod_{1}$}} & \mathbf{0}_{\frac{N}{9}} & \mathbf{H}_{p}\\
\mathbf{0}_{\frac{N}{9}} & \mathbf{0}_{\frac{N}{9}} & & & \mathbf{H}_{p} & \mathbf{0}_{\frac{N}{9}}\\
\end{array}
\right],
\label{eq:pcm_iraa_rc_r12}
\end{equation}
where $\mathbf{I}_{\frac{N}{9}}$, $\mathbf{H}_{2}$, $\mathbf{H}_{p}$ and
$\mathbf{0}_{\frac{N}{9}}$ are all $\frac{N}{9}\times \frac{N}{9}$ in dimension,
while $\prod_{1}$ is $\frac{N}{3}\times \frac{N}{3}$. The key point
to guarantee the full diversity property is the puncturing
procedure. Instead of puncturing $b$ parity bits we have punctured
$p$. The reason why puncturing $p$ instead of $b$ guarantees the
full diversity is due to the fact that the root-check structure of
the code is kept unchanged.

\subsubsection{IRAA Root-LDPC Rate $\frac{1}{3}$}

For the case of rate $1/3$ we considered the design done in
(\ref{eq:pcm_ira_rc_r13}) and we apply the constraints in
(\ref{eq:pcm_iraa}) to obtain the following parity-check matrix
\begin{equation}
\mathbf{H} =
\begin{bmatrix}
    \mathbf{H}_{u} & \vert & \mathbf{H}_{p} & \mathbf{0}\\
    \mathbf{0} & \vert & \prod_{1} & \mathbf{H}_{p}\\
\end{bmatrix}.
\label{eq:pcm_iraa_rc_r13}
\end{equation}
It must be noted that without puncturing the code rate is $1/5$.

\subsubsection{Pseudo-code for the IRA-PEG Root-LDPC Algorithm}
Initialization: A matrix of size $M \times N$ is created with the identity matrices $\mathbf{I}_{K}$ and parity matrices $\mathbf{H}_{p}$ in the positions shown in (\ref{eq:pcm_ira_rc_r12}), (\ref{eq:pcm_ira_rc_r13}), (\ref{eq:pcm_iraa_rc_r12}), (\ref{eq:pcm_iraa_rc_r13}) and zeros in all other positions. We define the indicator vectors ${\mathbf{z}}_{1}, \cdots, {\mathbf{z}}_{F}$ for the cases $R = \frac{1}{2}$, $R = \frac{1}{3}$ respectively as:
\begin{eqnarray}
{\mathbf{z}}_{1} & = & [\mathbf{0}_{1 \times \gamma}, \mathbf{1}_{1 \times \gamma}]^{T},\nonumber\\
{\mathbf{z}}_{2} & = & [\mathbf{1}_{1 \times \gamma}, \mathbf{0}_{1 \times \gamma}]^{T},\nonumber\\
\label{eq:ind_ira_f2}
\end{eqnarray}
\vspace{-.6cm}
\begin{eqnarray}
{\mathbf{z}}_{1} & = & [\mathbf{0}_{1 \times 2\chi}, \mathbf{1}_{1 \times \chi}, \mathbf{0}_{1 \times \chi}, \mathbf{1}_{1 \times \chi}, \mathbf{0}_{1 \times \chi}]^{T},\nonumber\\
{\mathbf{z}}_{2} & = & [\mathbf{1}_{1 \times \chi}, \mathbf{0}_{1 \times 4\chi}, \mathbf{1}_{1 \times \chi}]^{T},\nonumber\\
{\mathbf{z}}_{3} & = & [\mathbf{0}_{1 \times \chi}, \mathbf{1}_{1 \times \chi}, \mathbf{0}_{1 \times \chi}, \mathbf{1}_{1 \times \chi}, \mathbf{0}_{1 \times 2\chi}]^{T}\nonumber\\,
\label{eq:ind_ira_f3}
\end{eqnarray} where $\gamma = \frac{N}{2}$ for the case of IRA, while for IRAA design $\gamma = \frac{N}{4}$. We have $\chi = \frac{N}{9}$ for the case of IRA, while for IRAA design $\chi = \frac{N}{15}$. In addition, for rate $R = \frac{1}{2}$ under IRAA design ${\mathbf{z}}_{i} = [{\mathbf{z}}_{i}, \mathbf{0}_{4 \times \gamma}]$, while for rate $R = \frac{1}{3}$ under IRAA design ${\mathbf{z}}_{i} = [{\mathbf{z}}_{i}, \mathbf{0}_{6 \times \chi}]$.

These indicator vectors are modelled on that of the original PEG
algorithm \cite{hu.05}, indicating sub-matrices for which placement
is permitted, thus imposing the form of (\ref{eq:pcm_ira_rc_r12}),
(\ref{eq:pcm_ira_rc_r13}), (\ref{eq:pcm_iraa_rc_r12}),
(\ref{eq:pcm_iraa_rc_r13}). The degree sequence as defined for LDPC
codes must be altered to take into account the structure imposed by
Root-LDPC codes, namely, the identity matrices $\mathbf{I}_{K}$ and the
parity matrices $\mathbf{H}_{p}$, of (\ref{eq:pcm_ira_rc_r12}),
(\ref{eq:pcm_ira_rc_r13}), (\ref{eq:pcm_iraa_rc_r12}) and
(\ref{eq:pcm_iraa_rc_r13}). The pseudo-code for our proposed IRA-PEG
Root-LDPC algorithm is detailed in Algorithm \ref{alg:doprcpegalg1},
where the indicator vector $\mathbf{z}_i$ is taken from
(\ref{eq:ind_ira_f2}) and (\ref{eq:ind_ira_f3}) for constructing codes of rate
$R=\frac{1}{2}$, $R=\frac{1}{3}$ respectively.

\begin{algorithm}
 \caption{PEG Root-LDPC Algorithm}
 \label{alg:doprcpegalg1}
\algsetup{
linenosize=\small,
linenodelimiter=.
}
\begin{algorithmic}[1]
\FOR{$j = 1 : K$}
\FOR{$k = 0 : D_s(j)-1$}
\STATE Expand the PEG tree from the $j$\emph{-th} variable node to depth \emph{l} such that the tree contains all check nodes allowed by the indicator vector \textbf{or} the number of nodes in the tree does not increase with an expansion to the \emph{(l+1)-th} level.
\item[]
\STATE Place the edge connecting the $j$\emph{-th} variable node to a check node chosen randomly from the set of minimum weight nodes which were added to the sub-tree at the last tree expansion.
\ENDFOR
\ENDFOR
\end{algorithmic}
\end{algorithm}

\subsection{Controlled Doping Root-LDPC Codes Design}
Boutros in \cite{doping.boutros.11} proposed a controlled doping via high order Root-LDPC codes. Such Root-LDPC codes are able to guarantee full diversity for the parity check bits. First of all, we have made some modifications in the original doped Root-LDPC code parity-check matrix described in \cite{doping.boutros.11}.

\subsubsection{Controlled Doping Root-LDPC Codes $R = \frac{1}{2}$}
The modifications we have made was to take the advantages of easy encodability of IRA-based LDPC codes. Furthermore, a PEG-based design to improve the local girth of the generated LDPC codes was considered. Doping refers to the diversity achieved in the parity bits of the Root-LDPC graph, and when incidental is called uncontrolled. Controlled doping is used to intentionally improve the energy coefficient of information bits after solving parity bits. The energy coefficients relate the error rate achieved with the messages passed in decoding, in terms of the fading coefficients to which the code word is subjected \cite{Boutros.09.DPE}. Then, the parity bit should transmit a high-confidence message to a new information bit. Diversity population evolution (DPE) is an analytic method for studying the propagation of diversity in the graph during iterative decoding of a Root-LDPC code \cite{Boutros.09.DPE}. Uncontrolled doping corresponds to a DPE steady-state parameter $p_{\infty} = 7.82\%$ for a $C(3,6)$ regular Root-LDPC code \cite{doping.boutros.11}. Controlled doping can achieve a fraction $p_{\infty}$ as high as $100\%$. The sub-matrix (\ref{eq:pcm_ra_ira}) is modified as for the Root-LDPC-III code of \cite{doping.boutros.11} by introducing a smaller identity matrix for the parity bits. Therefore, the Root-LDPC code with $50\%$ of controlled doping, $H_{p}$ is redefined, to ensure a lower-triangular form and thus efficient encoding, as


\begin{equation}
\mathbf{H}_{p} = \begin{bmatrix}
\mathbf{I}_{\frac{N}{8}} & \mathbf{0}_{\frac{N}{8}} \\
\mathbf{P}_{\frac{N}{8}} & \mathbf{DD}_{\frac{N}{8}}
\end{bmatrix},
\label{eq:hp_doping}
\end{equation}
where $\mathbf{I}$ is an identity matrix, $\mathbf{0}$ is a null matrix, $\mathbf{P}$ is a permutation matrix with Hamming weight 1, $\mathbf{DD}$ is a dual diagonal matrix and all sub-matrices of $\mathbf{H}_{p}$ are $\frac{N}{8} \times \frac{N}{8}$ in dimension. Accordingly, the final parity-check matrix becomes
%
\begin{equation}
\mathbf{H}=
\begin{blockarray}{cccc}
1i & 2i & 1p & 2p \\
\begin{block}{[c|c|c|c]}
\mathbf{I}_{\frac{N}{4}} & \mathbf{H}_{2i} & \mathbf{0}_{\frac{N}{4}} & \begin{array}{cc}
\mathbf{I}_{\frac{N}{8}} & \mathbf{0}_{\frac{N}{8}} \\
\mathbf{P}_{2} & \mathbf{DD}_{\frac{N}{8}}\\
\end{array} \\
\BAhhline{||----||}
\mathbf{H}_{1i} & \mathbf{I}_{\frac{N}{4}} & \begin{array}{cc}
\mathbf{I}_{\frac{N}{8}} & \mathbf{0}_{\frac{N}{8}} \\
\mathbf{P}_{1} & \mathbf{DD}_{\frac{N}{8}}\\
\end{array} & \mathbf{0}_{\frac{N}{4}} \\
\end{block}
\end{blockarray},
\label{eq:pcm_rcdop_r12}
\end{equation}

where subscripts in $\mathbf{P}_{1}$ and $\mathbf{P}_{2}$ means that are distinct permutation sub-matrices with hamming weight 1. The sub-matrices $\mathbf{H}_{1i}$ and $\mathbf{H}_{2i}$ are in dimension $\frac{N}{4} \times \frac{N}{4}$. $\mathbf{P}_{1}$ and $\mathbf{P}_{2}$ are in dimension $\frac{N}{8} \times \frac{N}{8}$. The PEG algorithm will work through the sub-matrices $\mathbf{H}_{1i}$ and $\mathbf{H}_{2i}$.

\subsubsection{Controlled Doping Root-LDPC Codes $R = \frac{1}{3}$}
The parity-check matrix for the case of code rate $R = \frac{1}{3}$ has followed a similar design as for an IRA Root-LDPC code rate $R = \frac{1}{3}$ in  (\ref{eq:pcm_ira_rc_r13}). Therefore, the parity-check matrix for the proposed PEG controlled doping Root-LDPC code (PEG-CDRC LDPC) has the structure as presented in (\ref{eq:pcm_rcdop_r13}),
\begin{equation}
\mathbf{H} =
\begin{bmatrix}
\mathbf{H}_{2} & \mathbf{I} & \mathbf{0} & \vert & \mathbf{I} & \mathbf{0} & \mathbf{0} & \mathbf{0} & \mathbf{0} & \mathbf{0}\\
\mathbf{H}_{2} & \mathbf{0} & \mathbf{I}  & \vert & \mathbf{P}_{1} & \mathbf{DD} & \mathbf{0} & \mathbf{0} & \mathbf{0} & \mathbf{0}\\
\mathbf{I} & \mathbf{H}_{2} & \mathbf{0} & \vert & \mathbf{0} & \mathbf{0} & \mathbf{I} & \mathbf{0} & \mathbf{0} & \mathbf{0}\\
\mathbf{0} & \mathbf{H}_{2} & \mathbf{I} & \vert & \mathbf{0} & \mathbf{0} & \mathbf{P}_{2} & \mathbf{DD} & \mathbf{0} & \mathbf{0}\\
\mathbf{I} & \mathbf{0} & \mathbf{H}_{2} & \vert & \mathbf{0} & \mathbf{0} & \mathbf{0} & \mathbf{0} & \mathbf{I} & \mathbf{0}\\
\mathbf{0} & \mathbf{I} & \mathbf{H}_{2} & \vert & \mathbf{0} & \mathbf{0} & \mathbf{0} & \mathbf{0} & \mathbf{P}_{3} & \mathbf{DD}
\end{bmatrix},
\label{eq:pcm_rcdop_r13}
\end{equation}
where the subscripts of $\mathbf{P}_{i}$ in (\ref{eq:pcm_rcdop_r13}) means that are distinct permutation sub-matrices. The sub-matrices of eq. (\ref{eq:pcm_rcdop_r13}) are all $\frac{N}{9} \times \frac{N}{9}$ in dimension. In addition, the left hand side of (\ref{eq:pcm_rcdop_r13}) are connected to the information symbols while the right hand side are connected to the parity check bits.

\subsubsection{Controlled Doping Root-LDPC Codes $R = \frac{1}{4}$}
For the case of rate $\frac{1}{4}$ with $F = 4$, the Root-LDPC code structure with controlled doping is produced by a similar expansion of the parity-check matrix as from the graph for $F=2$ to the graph for $F=3$ described above. However, in addition we have adjusted the part of the matrix associated with the parity bits to account for the dimension requirements of the Root-LDPC structure at this rate, where each of the four $\mathbf{H}_p$ matrices have dimension $\frac{3N}{16} \times \frac{3N}{16}$ and as such have been adjusted to take the structure of (\ref{eq:Hp_doping_r14}).

\begin{equation}
\mathbf{H}_{p} = \begin{bmatrix}
\mathbf{I}_{\frac{N}{16}} & \mathbf{0}_{\frac{N}{16}} & \mathbf{0}_{\frac{N}{16}} \\
\mathbf{P}_{\frac{N}{16}} & \mathbf{I}_{\frac{N}{16}} & \mathbf{0}_{\frac{N}{16}} \\
\mathbf{P}_{\frac{N}{16}} & \mathbf{P}_{\frac{N}{16}} & \mathbf{DD}_{\frac{N}{8}}
\end{bmatrix},
\label{eq:Hp_doping_r14}
\end{equation}

and the matrices $\mathbf{I}$, $\mathbf{P}$ and $\mathbf{DD}$ are as defined previously for the cases of $F=2$ and $F=3$. Note in particular that the permutation matrices $\mathbf{P}$ each have distinct cyclic shifts.

\subsection{Proposed Design Algorithm}
Here, we introduce some definitions and a specific notation. Then, the construction for the proposed codes is carried out by the pseudo-code previously introduced in Algorithm \ref{alg:doprcpegalg1}, along with appropriate initialization and using the indicator vectors defined in the following. In this work, the scenarios of a block-fading channel with $F = 2$, $F = 3$ and $F = 4$ are considered. In extending to a greater number of fadings, $F>4$, the general structure presented is maintained.

\subsubsection{Pseudo-code for the PEG-CDRC LDPC Algorithm}
Initialization: A matrix of size $M \times N$ is created with the identity matrices $\mathbf{I}$, dual diagonal matrices $\mathbf{DD}$ and parity matrices $\mathbf{P}_{i}$ in the correct positions and zeros in all other positions, as shown  in (\ref{eq:pcm_rcdop_r12}) and (\ref{eq:pcm_rcdop_r13}), and similarly for the code for the $F=3$ channel, using instead $\mathbf{H}_p$ of (\ref{eq:Hp_doping_r14}).
We define the indicator vectors ${\mathbf{z}}_{1}, \cdots, {\mathbf{z}}_{F}$ for the cases $R = \frac{1}{2}$, $R = \frac{1}{3}$ and $R = \frac{1}{4}$ respectively as:
\begin{eqnarray}
{\mathbf{z}}_{1} & = & [\mathbf{0}_{1 \times \gamma}, \mathbf{1}_{1 \times \gamma}]^{T},\nonumber\\
{\mathbf{z}}_{2} & = & [\mathbf{1}_{1 \times \gamma}, \mathbf{0}_{1 \times \gamma}]^{T},\nonumber\\
\label{eq:inddopf2}
\end{eqnarray}
\vspace{-.6cm}
\begin{eqnarray}
{\mathbf{z}}_{1} & = & [\mathbf{0}_{1 \times 2\chi}, \mathbf{1}_{1 \times \chi}, \mathbf{0}_{1 \times \chi}, \mathbf{1}_{1 \times \chi}, \mathbf{0}_{1 \times \chi}]^{T},\nonumber\\
{\mathbf{z}}_{2} & = & [\mathbf{1}_{1 \times \chi}, \mathbf{0}_{1 \times 4\chi}, \mathbf{1}_{1 \times \chi}]^{T},\nonumber\\
{\mathbf{z}}_{3} & = & [\mathbf{0}_{1 \times \chi}, \mathbf{1}_{1 \times \chi}, \mathbf{0}_{1 \times \chi}, \mathbf{1}_{1 \times \chi}, \mathbf{0}_{1 \times 2\chi}]^{T},\nonumber\\
\label{eq:inddopf3}
\end{eqnarray}
\vspace{-.6cm}
\begin{eqnarray}
{\mathbf{z}}_{1} & = & [\mathbf{1}_{1 \times 3\zeta}, \mathbf{0}_{1 \times 5\zeta}, \mathbf{1}_{1 \times \zeta}, \mathbf{0}_{1 \times \zeta}, \mathbf{1}_{1 \times \zeta}, \mathbf{0}_{1 \times \zeta}]^{T},\nonumber\\
{\mathbf{z}}_{2} & = & [\mathbf{0}_{1 \times 3\zeta}, \mathbf{1}_{1 \times 3\zeta}, \mathbf{0}_{1 \times 2\zeta}, \mathbf{1}_{1 \times 2\zeta}, \mathbf{0}_{1 \times 2\zeta}]^{T},\nonumber\\
{\mathbf{z}}_{3} & = & [\mathbf{1}_{1 \times \zeta}, \mathbf{0}_{1 \times \zeta}, \mathbf{1}_{1 \times \zeta}, \mathbf{0}_{1 \times 2\zeta}, \mathbf{1}_{1 \times 3\zeta}, \mathbf{0}_{1 \times 4\zeta}]^{T},\nonumber\\
{\mathbf{z}}_{4} & = & [\mathbf{1}_{1 \times \zeta}, \mathbf{0}_{1 \times 2\zeta}, \mathbf{1}_{1 \times \zeta}, \mathbf{0}_{1 \times 5\zeta}, \mathbf{1}_{1 \times 3\zeta}]^{T},\nonumber\\
\label{eq:inddopf4}
\end{eqnarray} where $\gamma = \frac{N}{4}$, $\chi = \frac{N}{9}$ and $\zeta = \frac{N}{16}$.

These indicator vectors are modelled on the basis of the original PEG algorithm \cite{hu.05}, indicating sub-matrices for which placement is permitted, thus imposing the Controlled Doping Root-LDPC form. The degree sequence as defined for LDPC codes must be altered to take into account the structure imposed by Root-LDPC codes, namely, the identity matrices $\mathbf{I}$, the permutation matrices $\mathbf{P}_{i}$, the dual diagonal matrices $\mathbf{DD}$ and the parity matrices $\mathbf{H}_{i}$, of (\ref{eq:pcm_rcdop_r12}), (\ref{eq:pcm_rcdop_r13}) and the multiple uses of (\ref{eq:Hp_doping_r14}). The  proposed CDRC-LDPC construction algorithm is then implemented using Algorithm \ref{alg:doprcpegalg1}, with the parity-check matrix suitably initialised with matrices $\mathbf{I}$, $\mathbf{DD}$ and distinct random permutation matrices $\mathbf{P}_i$ in the appropriate positions. The indicator vectors $\mathbf{z}_i$ are taken from (\ref{eq:inddopf2}), (\ref{eq:inddopf3}) and (\ref{eq:inddopf4}) for constructing codes of rate $R=\frac{1}{2}$, $R=\frac{1}{3}$ and $R=\frac{1}{4}$ respectively.

\section{Discussion}
In this section we analyse the advantages and disadvantages of different types of PEG-based Root-LDPC codes discussed in the previous sections.

In terms of performance the PEG-based Root-LDPC codes are able to get closer to the outage curve than their counterpart Root-LDPC codes. However, the complexity of encoding standard PEG-based Root-LDPC codes can be prohibitive for some hardware implementations.

The Quasi-Cyclic PEG-based Root-LDPC codes have the advantage of performing better than Quasi-Cyclic Root-LDPC codes and both codes require low memory to store the parity-check matrix. Moreover, Quasi-Cyclic based LDPC codes can be encoded by using simple shift registers.

RA PEG-based Root-LDPC codes have the advantage of being simple to encode and also simple to design the parity-check matrix. Furthermore, the parity part of an RA-based parity-check matrix is a dual diagonal which is straightforward to obtain the generator matrix \cite[pp. 267-279]{ryanbook}. Such codes perform very close to the channel capacity which is usually upper-bounded by the outage curve. In addition, RA-based LDPC codes can provide: low complexity to encode, simplicity on the design of the parity-check matrix and low memory is required to store them. On the other hand, the main limitation of RA-based codes is the code rate, which cannot be higher than $\frac{1}{2}$.

In the case of Unstructured Full Diversity LDPC codes, they draw an important path in terms of designing the parity-check matrix which avoids the constraints that must be imposed to produce root-check based LDPC codes. Nonetheless, they require a more complex encoding process which is the same complexity as the case of Random LDPC codes.

As discussed previously, the PEG Controlled Doping Root-LDPC codes are able to guarantee full diversity for the parity check bits. These LDPC codes are relevant for the case of IDD in MIMO systems. The results presented in \cite{wcnc.14} demonstrate how useful are PEG-CDRC LDPC codes for MIMO systems in a block-fading channel. In addition, our proposed PEG-CDRC LDPC codes have the advantage of being RA-based encodable which are simple to encode and the parity-check matrix is easily designed.

\section{Simulations} \label{sec:simul}

The performance of the proposed PEG-based Root-LDPC codes for block-fading
channels with $F = 2$, $F = 3$ and $F = 4$ independent fading blocks is
analysed. The block length of the codes for rates $R = \frac{1}{2}$ and $R =
\frac{1}{4}$ is $N = 1024$ while for rate $R = \frac{1}{3}$ the block length is
$N = 900$. Iterative message passing is employed at the decoder with a maximum
of $5$ iterations for rate $R = \frac{1}{2}$ and for rates $R = \frac{1}{3}$
and $R = \frac{1}{4}$ a maximum of $20$ iterations were used. The Gaussian
outage limit in (\ref{eq:outage}) is drawn in dashed line in each figure for
reference.

In Fig. 4 we compare the FER performance among the proposed PEG-CDRC LDPC
codes, IRA PEG Root-LDPC code, IRAA PEG Root-LDPC codes, QC-PEG-Root-LDPC and
PEG-Root-LDPC, Random Root-LDPC and PEG based LDPC \cite{hu.05} codes, all for
$R=\frac{1}{2}$. From the results, it can be noted  that the proposed PEG-CDRC
LDPC code, IRAA-PEG Root-LDPC code and PEG-Root-LDPC code achieve the same FER
performance. Moreover, note that all root-check-based codes are able to achieve
the full diversity order of the channel, while (non root-check based) PEG LDPC
codes fail to achieve full diversity. The PEG-based Root-LDPC codes outperform
the PEG LDPC code by $7.5$dB at a FER between $10^{-2}$ and $10^{-3}$.

In Fig. 5 we compare the FER performance between the proposed PEG-CDRC LDPC,
QC-PEG-Root-LDPC, IRA PEG Root-LDPC code, IRAA PEG Root-LDPC codes, QC-PEG LDPC
codes and PEG-root-LDPC code, all for $R=\frac{1}{3}$. From the results, it can
be seen that the best performance is achieved by the proposed Quasi-Cyclic PEG
Root-LDPC code. IRA-PEG Root-LDPC and IRAA-PEG Root-LDPC have in average the
same performance in terms of FER. The PEG-CDRC LDPC code is performing
marginally worse than IRA and IRAA PEG root-check based LDPC codes. It was
required to sacrifice the FER performance of the proposed PEG-CDRC LDPC codes
to guarantee the full diversity of the parity check bits. Moreover, note that
the proposed CDRC-LDPC code outperforms the QC-PEG LDPC code consistently
across the range of FER considered, with an improvement of $2$dB below a FER of
$10^{-3}$. The proposed QC-PEG-Root-LDPC code outperforms the QC-PEG LDPC code
by about $3.5$dB, also between a FER of $10^{-3}$ and $10^{-4}$.

In Fig. 6 we compared the FER performance between the proposed PEG-CDRC LDPC,
QC-PEG-Root-LDPC codes, IRA-PEG-Root-LDPC and QC-PEG LDPC codes all for
$R=\frac{1}{4}$. The codeword length is $N=1024$ bits. From the results, it can
be noted that the proposed PEG-CDRC LDPC code outperforms the QC-PEG LDPC code
by about $1.5$dB while the proposed QC-PEG-Root-LDPC code outperforms the
QC-PEG LDPC code by about $2.5$dB. In addition, note that only the PEG-based
Root-LDPC codes are able to achieve the full diversity order of the channel.
For the PEG-CDRC LDPC code, the FER of the whole code word is also included at
both $20$ and $100$ maximum decoder iterations. Note that the whole code word
error rate at $20$ maximum decoder iterations is dominated by the
unsatisfactory performance of the parity bits, but at the higher maximum number
of decoder iterations the whole code word FER has converged to that of the
information bits, demonstrating that the controlled doping has had the desired
effect. Recall that the doping used leads to $p_{\infty}=100\%$, which is the
percentage of variable nodes corrected after an arbitrarily large number of
decoder iterations, and so this behaviour is expected from the PEG-CDRC code.
Finally note that both IRA-PEG-Root-LDPC and PEG-CDRC codes exhibit a loss in
performance with respect to the QC-PEG-Root-LDPC code. This results from the
combined repeat-accumulate and Root-LDPC structures found in the graphs of
those codes, which offer a reduction in encoding complexity and
diversity-achieving performance at the expense of reduced coding gain.

Fig. 7 shows the average number of iterations required by the proposed PEG-CDRC
LDPC codes, IRA PEG Root-LDPC code, IRAA PEG Root-LDPC codes, PEG-Root-LDPC,
Random Root-LDPC and PEG based LDPC \cite{hu.05} codes, all for
$R=\frac{1}{2}$. The decoder was operated to a maximum of $5$ iterations and
with the zero syndrome stopping criterion in place. Other decoding algoriths
can also be considered \cite{vfap}. For the entire SNR region, in average, we
can observe that the proposed PEG root-check based LDPC codes require less
decoding iterations than standard PEG LDPC code. It must be mentioned that for
medium to high SNR the average required number of iterations is less than $2$
iterations. The average number of iterations, less than $2$ at medium to high
SNR, corroborates with hardware friendly capabilities of structured LDPC codes
\cite{salehi.10}.

\section{Conclusion}
Novel PEG-based algorithms have been proposed to design Controlled Doping Root-LDPC codes, IRA Root-LDPC codes, IRAA Root-LDPC codes and Quasi-Cyclic Root-LDPC codes for $F \geq 2$ fading blocks. Based on simulations, the proposed methods were compared to non Root-LDPC codes. The results demonstrate that the root-check based LDPC codes generated by our proposed algorithm outperform standard LDPC codes. Furthermore, for the case of rate $R = \frac{1}{2}$ the PEG-based Root-LDPC codes outperform the PEG LDPC code by about $7.5$dB. As mentioned before, the proposed PEG-CDRC LDPC codes are RA based LDPC codes which are simple to encode and the parity-check matrix can be easily designed.



\section*{Competing interests}
  The authors declare that they have no competing interests.


\section*{Acknowledgements}
  This work was partially supported by PNPD/CAPES and CNPq (Brazil), under grant 237676/2012-5.




\section*{Figures}

\begin{figure}[!h]
\caption{Parity-check matrix unstructured general case. Parity-check matrix for
the general case.} \label{fig:PCM_gen}
\end{figure}

\begin{figure}[!h]
\caption{RA code block diagram. A systematic repeat-accumulate code block
diagram, where $K$ is the number of information bits and $p$ denotes the parity
bits.} \label{fig:ra_diagram}
\end{figure}

\begin{figure}[!h]
\caption{IRAA code block diagram.  A systematic irregular repeat-accumulate and
accumulate code block diagram. Where $K$ are the information bits, $b$ and $p$
are the parity bits.} \label{fig:iraa_diagram}
\end{figure}

\begin{figure}[!h]
\caption{FER performance $F = 2$.  FER performance for the PEG-CDRC LDPC,
IRA-PEG Root-LDPC, IRAA-PEG Root-LDPC, PEG-Root-LDPC, Random Root-LDPC and PEG
LDPC codes over a block-fading channel with $F = 2$ and $N = 1024$ bits. The
maximum number of iterations is $5$.} \label{fig:fig_r12}
\end{figure}

\begin{figure}[!h]
\caption{FER performance $F = 3$. FER performance for the CDRC-LDPC,
QC-PEG-Root-LDPC , IRA-PEG Root-LDPC, IRAA-PEG Root-LDPC and QC-PEG LDPC codes
over a block-fading channel with $F = 3$ and $N = 900$ bits. The maximum number
of iterations is $20$.} \label{fig:fig_r13}
\end{figure}

\begin{figure}[!h]
\caption{FER performance $F = 4$. FER performance for the PEG-CDRC LDPC,
QC-PEG-Root-LDPC and QC-PEG LDPC codes over a block-fading channel with $F = 4$
and $N = 1024$ bits. The maximum number of iterations is $20$.}
\label{fig:fig_r14}
\end{figure}

\begin{figure}[!h]
\caption{Iterations performance comparison for $F = 2$. Average number of
required iterations for the proposed PEG-CDRC LDPC codes, IRA PEG Root-LDPC
code, IRAA PEG Root-LDPC codes, PEG-Root-LDPC, Random Root-LDPC and PEG based
LDPC codes with codeword length $N = 1024$ bits over a block-fading channel
with $F = 2$. Maximum number of iterations $5$.} \label{fig:fig_iter_r12}
\end{figure}



\end{document}